\documentclass[sigconf]{acmart}
\AtBeginDocument{%
  }

\copyrightyear{2025}
\acmYear{2025}
\setcopyright{rightsretained}
\acmConference[IUI Companion '25]{30th International Conference on Intelligent User Interfaces Companion}{March 24--27, 2025}{Cagliari, Italy}
\acmBooktitle{30th International Conference on Intelligent User Interfaces Companion (IUI Companion '25), March 24--27, 2025, Cagliari, Italy}\acmDOI{10.1145/3708557.3716148}
\acmISBN{979-8-4007-1409-2/25/03}

\usepackage{csquotes}
\usepackage{enumitem}

\begin{document}

\title[Interview AI-ssistant]{Interview AI-ssistant: Designing for Real-Time Human-AI Collaboration in Interview Preparation and Execution}

\author{Zhe Liu}
\email{zheliu92@cs.ubc.ca}
\orcid{0000-0002-1904-9045}
\affiliation{%
  \institution{University of British Columbia}
  \city{Vancouver}
  \state{BC}
  \country{Canada}
}


\begin{abstract}
Recent advances in large language models (LLMs) offer unprecedented opportunities to enhance human-AI collaboration in qualitative research methods, including interviews. While interviews are highly valued for gathering deep, contextualized insights, interviewers often face significant cognitive challenges, such as real-time information processing, question adaptation, and rapport maintenance. My doctoral research introduces \textit{Interview AI-ssistant}, a system designed for real-time interviewer-AI collaboration during both the preparation and execution phases. Through four interconnected studies, this research investigates the design of effective human-AI collaboration in interviewing contexts, beginning with a formative study of interviewers’ needs, followed by a prototype development study focused on AI-assisted interview preparation, an experimental evaluation of real-time AI assistance during interviews, and a field study deploying the system in a real-world research setting. Beyond informing practical implementations of intelligent interview support systems, this work contributes to the Intelligent User Interfaces (IUI) community by advancing the understanding of human-AI collaborative interfaces in complex social tasks and establishing design guidelines for AI-enhanced qualitative research tools.
\end{abstract}



\begin{teaserfigure}
    \centering
    \begin{minipage}{0.32\linewidth}
        \centering
        \includegraphics[width=\linewidth]{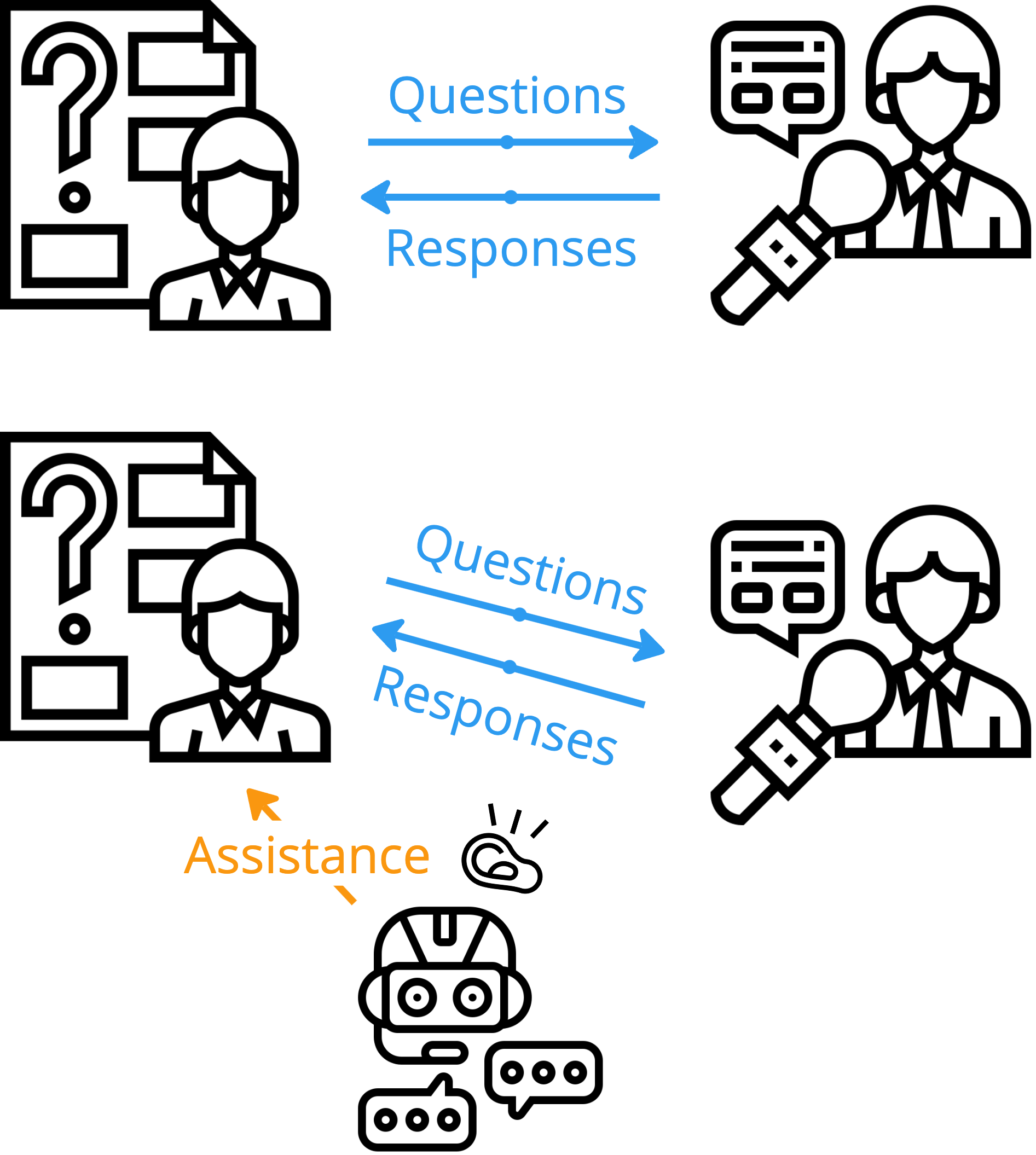}
        \caption{Traditional interview (top) and AI-assisted interview (bottom) scenarios}
        \label{fig:figure1}
    \end{minipage}
    \hfill
    \begin{minipage}{0.62\linewidth}
        \centering
        \includegraphics[width=\linewidth]{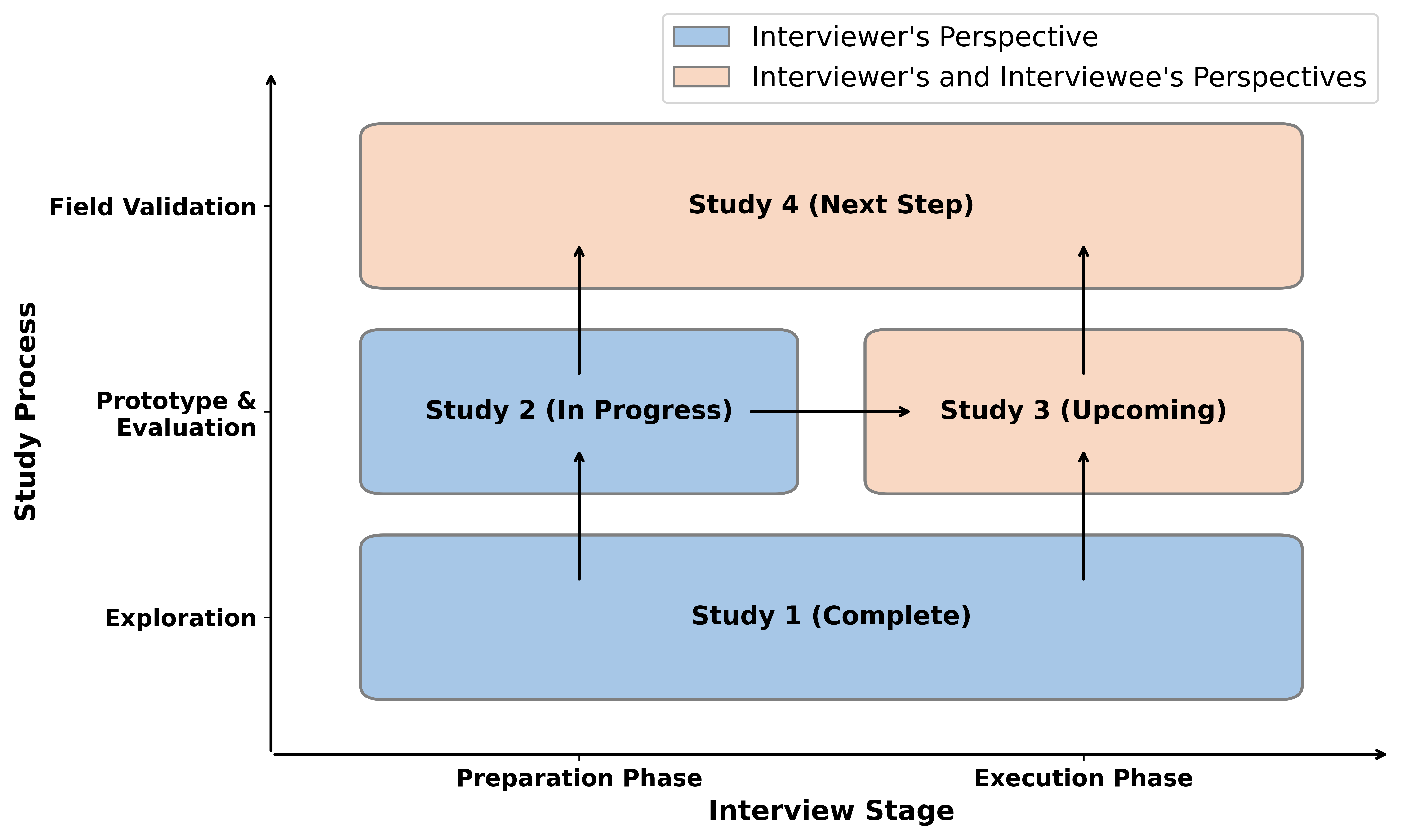}
        \caption{Overview of the research plan (four studies covering both the interview preparation and execution phases)}
        \label{fig:figure2}
    \end{minipage}
\end{teaserfigure}

\received{January 2025}
\received[accepted]{January 2025}

\maketitle

\section{INTRODUCTION \& MOTIVATION}
Recent advances in large language models (LLMs) have opened new opportunities for human-AI collaboration in qualitative research methods \cite{xiao2020if, xiao2021let, bulygin2022conversational}. Among these methods, interviews stand out as a flexible yet systematic approach to collecting rich, nuanced data \cite{morris2015practical}. However, this flexibility places significant cognitive demands on interviewers, particularly novices \cite{gesch2015reflecting}, who must simultaneously process complex verbal and nonverbal cues \cite{morris2015practical}, adapt questions strategically to elicit deeper insights \cite{houtkoop1996probing}, and sustain rapport through responsive dialogue \cite{morris2015practical}. While fully automated AI-driven solutions often fall short of accommodating the flexibility and nuance that interviews require \cite{shneiderman2022human}, real-time AI interview assistants hold the potential to augment—rather than replace—human judgment \cite{lubars2019ask} by scaffolding these cognitive processes and addressing key challenges through interviewer-AI collaboration.

Yet, despite the opportunities presented by human-AI collaboration, existing research has predominantly focused on \textit{asynchronous} tasks, such as post-hoc support for qualitative data analysis \cite{jiang2021supporting, feuston2021putting}. As illustrated in Figure \ref{fig:figure1}, which contrasts traditional interview scenarios with AI-assisted ones, the design of \textit{synchronous}, in-the-moment collaborative interfaces—especially for interview support—remains largely underexplored, limiting our understanding of how AI can dynamically assist during active interactions.

Therefore, my doctoral research investigates \textit{Interview AI-ssistant}, a real-time AI-integrated system for interviewer-AI collaboration, facilitating \textit{pre-interview} preparation and providing real-time assistance \textit{during interviews}. Beyond empowering academic researchers and industry practitioners in qualitative research, this work aims to contribute to the IUI community by advancing the understanding of intelligent collaborative interfaces and establishing guidelines for AI-enhanced qualitative research tools. Ultimately, it seeks to foster interdisciplinary dialogue for more insightful, efficient, and ethically sound human-AI collaboration in social science research.

\section{GOAL \& RESEARCH QUESTIONS:} 
The primary goal of this doctoral research is to develop and evaluate \textit{Interview AI-ssistant}, a real-time AI-integrated system designed to facilitate interviewer-AI collaboration. In addition to enhancing both the pre-interview preparation phase and the interview execution phase, this research will also carefully consider the perspectives of all stakeholders involved in the AI-assisted interview context, namely both interviewers and interviewees. To guide this research endeavor, we propose the following three research questions (RQs):

\begin{itemize}[leftmargin=1em]
    \item \textbf{RQ1}: How can AI technologies be effectively integrated to enhance both pre-interview preparation and real-time facilitation during interviews, and what design strategies can be implemented to mitigate the limitations of current AI technologies?
    \item \textbf{RQ2}: How do \textit{interviewers} perceive the real-time interviewer-AI collaboration, and what impact does this collaboration have on the interview quality and effectiveness?
    \item \textbf{RQ3}: How do \textit{interviewees} perceive and respond to AI-assisted interviews, and what ethical considerations arise from this collaborative approach?
\end{itemize}

\section{RELATED WORK}
\subsection{AI Advances in Supporting Qualitative Research}
An increasing number of publications explore AI's role in supporting qualitative research, particularly in structured data collection \cite{xiao2023supporting} and exploratory qualitative analysis \cite{feuston2021putting}, evolving alongside advancements in AI technologies. Beyond AI’s ability to automate transcription\footnote{e.g., https://openai.com/research/whisper}, several studies have examined the use of chatbots\footnote{Referring to AI-driven conversational agents designed to interact with humans \cite{adamopoulou2020overview}} for collecting structured qualitative data in surveys with open-ended questions. By fostering social interactions \cite{xiao2020tell} and practicing active listening \cite{xiao2020if}, these chatbots have been shown to enhance participant engagement and elicit higher-quality responses compared to traditional online questionnaires. However, since these chatbots typically rely on predefined conversational capabilities and scripted questions, they lack the adaptability required for semi-structured interviews \cite{bulygin2022conversational}.

Researchers have also explored AI’s potential to comprehend qualitative data and identify patterns \cite{feuston2021putting, jiang2021supporting}, performing tasks such as theme suggestion and automatic coding \cite{kuang2024enhancing}. Despite progress, current AI-generated outputs often fail to meet human-level standards, functioning more as preliminary processing tools \cite{feuston2021putting}. Some studies have investigated AI’s utility in more closed-ended tasks, including text-based deductive coding \cite{xiao2023supporting} and audio or video-based emotion detection \cite{pfeifer1988artificial, graterol2021emotion}, with positive results. Nonetheless, these prototypes face challenges with transparency and explainability. Feuston and Brubaker further examined qualitative scholars' perspectives on collaborating with AI for open-ended inductive analysis, though practical design implementations of their findings remain limited \cite{feuston2021putting}. Since most existing systems are designed for post-collection qualitative analysis, they primarily offer AI assistance in an “asynchronous” manner—providing suggestions \cite{xiao2023supporting} or interpretations \cite{feuston2021putting} of the analysis for researchers to review and adopt later. However, the real-time challenges faced by interviewers during semi-structured interviews demand “synchronous” support throughout the interview process \cite{morris2015practical}, highlighting the inadequacy of existing asynchronous tools. Moreover, user preferences and interactions with real-time AI assistance in qualitative methodology remain largely unexplored.

\subsection{Designing for Human-AI Collaboration}

Driven by the goal of preserving human autonomy—defined as maintaining human agency and decision-making control \cite{jiang2021supporting}—there is growing interest in human-AI collaboration across domains ranging from problem-solving to creative endeavors. While traditional problem-solving tasks, such as mathematical \cite{memmert2022complex} or technical challenges \cite{jonsson2022cracking}, often involve well-defined objectives and a range of effective solutions, creative domains are inherently more open-ended and less predictable, aligning closely with the dynamic and adaptive nature of interviews. Studies have demonstrated how AI can enhance human creativity in areas such as writing \cite{biermann2022writers}, drawing \cite{oh2018lead}, and music composition \cite{louie2020novice}, helping users achieve higher-quality outcomes. However, these collaborations remain predominantly asynchronous, with concerns about AI’s controllability, transparency, and impact on the originality and ownership of creative outputs \cite{oh2018lead}.

In multi-stakeholder contexts, such as interviews \cite{morris2015practical} or doctor-patient interactions \cite{lorenzini2023artificial}, research has emphasized the importance of transparency, explainability, and striking a balance between AI automation and human control \cite{amershi2019guidelines}. Real-time AI support in medical decision-making has shown potential as a mediator \cite{bulla2020review}, offering benefits such as improved decision-making and shared understanding. These include AI’s roles in pre-diagnosis data collection \cite{bulla2020review}, real-time shared decision-making \cite{aminololama2019doctor}, and post-diagnosis documentation \cite{kocaballi2020envisioning}. However, it also highlights challenges related to privacy, trustworthiness, and user preferences \cite{lorenzini2023artificial}. Parallels between medical and interview contexts, including knowledge imbalance, information asymmetry, and communication dynamics \cite{riley2003exploring}, underscore the relevance of lessons from AI-assisted medical applications. 

Despite these parallels, interviews present unique challenges that make real-time AI-driven assistance particularly complex and they are underexplored. Beyond adapting to dynamic conversational flows and shifting responses, AI assistance is expected to support interviewers in eliciting deep, rich, and authentic qualitative data through context-sensitive probing. These demands require flexibility and interactivity that existing tools—largely focused on asynchronous support or rigidly structured data collection—fail to address. Building on prior research that emphasizes augmenting rather than automating human practices \cite{feuston2021putting}, this work seeks to bridge the gap between asynchronous tools and interactive AI, exploring adaptive, synchronous AI systems tailored specifically for interviews, with the aim of enhancing interviewer experiences and improving the quality of qualitative data collection.

\section{RESEARCH PLAN}
My doctoral research investigates interviewer-AI collaboration through four interconnected studies that span both the preparation and execution phases of interviews (see Figure \ref{fig:figure2}). \textbf{Study 1}, recently accepted to Computer-Supported Cooperative Work and Social Computing (CSCW) 2025 \cite{zhe2025cscw} explores the current practices and needs of interviewers in relation to AI assistance. Based on these insights, \textbf{Study 2} focuses on developing and evaluating an AI-assisted interview preparation system, targeting efficiency and skill development. Building on the findings from Study 2, \textbf{Study 3} assesses the effectiveness of real-time AI support during interviews through controlled experiments, capturing both interviewer and interviewee perspectives. Finally, \textbf{Study 4} applies the refined AI system in a field study to examine its real-world impact on research practice. This progressive approach ensures a deep understanding of interviewers’ nuanced requirements and enables the iterative development of AI tools that directly support human-AI collaborative interviewing in practice.

\subsection*{Study 1 (Complete): Exploratory Interviews with Interviewers Across the Expertise Spectrum}

\textit{Methods.\ } To understand interviewers’ perspectives on AI assistance, I conducted an exploratory formative study with 16 researchers possessing varying levels of interviewing expertise, ranging from novice to expert. This selection allowed for a broader understanding of the diverse needs and expectations of interviewers. The study included semi-structured interviews and an inductive thematic analysis to examine participants’ visions and expectations for AI-driven interview assistants, uncovering both the potential and challenges in real-time assistance.

\textit{Results.\ } This exploratory study revealed two key findings. First, interviewers across expertise levels expect real-time AI assistance to serve dual purposes: supporting research objectives and facilitating interpersonal communication, though they expressed concerns about its potential impact on long-term skill development. Second, the study identified three critical contextual factors that shape human-AI collaboration in interview settings: (1) time constraints during live interactions, (2) the need for multimodal collaboration, and (3) the importance of managing interviewee-AI interactions within the triangular dynamic among interviewers, interviewees, and AI systems. The study also uncovered several design challenges specific to developing expertise-centric AI assistant systems for interviewers, particularly in balancing AI support with the cultivation of interviewer expertise \cite{zhe2025cscw}.

\subsection*{Study 2 (In Progress): Developing and Evaluating an AI-Assisted Interview Practice System}

Building on insights from Study 1, I will employ an iterative, user-centered design approach to develop and evaluate an AI-integrated system prototype for improving interview practice efficiency. The study will involve 20 participants with diverse backgrounds and expertise levels from academic and professional sectors, using mixed methods including surveys, system logs, and post-evaluation interviews. Participants will be divided into two conditions: one with real-time AI assistance and one with non-real-time AI assistance, with the latter serving as the baseline. Through this evaluation, we aim to understand the effectiveness of AI-assisted interview practice, particularly how real-time assistance impacts practice efficiency and skill development across different expertise levels. Additionally, we will assess the users’ perceptions of the AI system’s usability and its ability to support the actual conduct of interviews. The findings will reveal interaction patterns and design considerations for interviewer-AI collaboration, informing the development of real-time AI assistance features for Study 3. This approach will inform the refinement of the system’s adaptability and utility for interviewers with varying levels of expertise, from novice to expert.

\subsection*{Study 3 (Upcoming): Experimental Evaluation of Real-Time AI-Enhanced Interviewing}

Following the development of the prototype in Study 2, this study will assess the effectiveness of real-time AI assistance through a controlled experimental design, where 20 pairs of participants will conduct parallel interviews with and without the \textit{Interview AI-ssistant}. Interview quality will be evaluated using established metrics, including question depth, topic coverage, and follow-up effectiveness. In addition to the perspectives of interviewers, interviewees will also provide feedback, a critical addition not explored in previous studies. Understanding the views of both parties is vital for capturing the full impact of AI assistance on the interview process, especially as it informs the real-world deployment in Study 4. A mixed-methods analysis will combine these quantitative metrics with post-interview debriefings from both interviewers and interviewees, offering a comprehensive view of the perceived influence of real-time AI support. This evaluation aims to understand how AI affects interview quality, collaboration patterns, and ethical considerations such as trust and privacy. The findings will directly inform design refinements for the real-world deployment of AI assistance in research settings in Study 4.

\subsection*{Study 4 (Next Step): Field Deployment of AI-Enhanced Interviewing in Research Practice}

Building upon findings from Study 3, this study will deploy and iterate on the \textit{Interview AI-ssistant} prototype in a small-scale field study with 3-5 researchers conducting their own interviews. We will observe how the AI-enhanced system adapts to real-world research practices, collecting both qualitative and quantitative data, including system logs, interview quality assessments (e.g., question depth, follow-up effectiveness), and post-study interviews to gather feedback on challenges and perceived impact. This study will examine the influence of real-time AI assistance on interview quality and efficiency, as well as its integration into researchers’ workflows in dynamic, unpredictable contexts. By moving beyond controlled environments, this study will provide valuable insights into AI’s practical applications and limitations, refining the system based on real user experiences and enhancing our understanding of its potential in shaping future qualitative research.

\section{LONG TERM GOAL}
My long-term goal is to advance AI-assisted interviewing tools that transform academic and industry research practices. By collaborating with researchers and practitioners, I aim to design and implement intelligent systems that enhance real-time human-AI collaboration in qualitative research, enabling richer and more efficient data collection. While my PhD focuses on foundational research and prototype development, I envision extending this work to address practical challenges such as domain-specific customizations, and ethical considerations. Achieving this vision will require close engagement with diverse stakeholders to ensure these tools are not only innovative but also adaptable and impactful across various research contexts.

\begin{acks}
I would like to thank my supervisor, Dr. Joanna McGrenere, for her thoughtful feedback and unwavering encouragement throughout my research. I appreciate Dr. Vered Shwartz’s technical expertise in natural language processing, which has greatly deepened my understanding of AI technologies. I am grateful to Dr. Jiamin Dai for her insights into qualitative methodology, which were instrumental in shaping the research direction. Finally, my gratitude goes to all the participants for their invaluable perspectives.
\end{acks}

\bibliographystyle{ACM-Reference-Format}
\bibliography{sample-base}

\appendix

\end{document}